\documentclass[linenumbers]{aastex631}
\usepackage{multirow}
\usepackage{soul}

\graphicspath{{./}{figures/}}
\begin{document}

\title{Generation and Life Cycle of Solar Spicules}

\author{Hamid Saleem}
\affiliation{Theoretical Research Institute, Pakistan Academy of Sciences, 3-Constitution Avenue, G-5/2, Islamabad 44000, Pakistan.}
\affiliation{Department of Physics, School of Natural Sciences (SNS), National University of Sciences and Technology (NUST), H-12, Islamabad 44000, Pakistan}
\affiliation{Space and Astrophysical Research Laboratory (SARL),
National Center of GIS and Space Applications,
Islamabad 44000, Pakistan}

\author{Zain H. Saleem}
\affiliation{Argonne National Laboratory,
9700 S. Cass Ave,
Lemont, IL 60439, USA.}



\begin{abstract}
Physical mechanism for the creation of solar spicules is proposed with three stages of their life cycle. It is assumed that at stage-I, the density hump is formed locally in the xy-plane in lower chromosphere in the presence of temperature gradients of electrons and ions along z-axis (the vertical direction). In this region, the density structure of quasi-neutral $(n_i\simeq n_e=n)$ plasma after taking birth is accelerated in the vertical direction due to the thermal force ${\bf F}_{th} \propto \nabla n(x,y,t) \times (\nabla T_e + \nabla T_i)$. The exact time-dependent  analytical solution of two fluid plasma equations is presented assuming that density is maximum at the center and decays away from it gradually. The two dimensional (2D) density structure is created as a step function $H(t)$ in time at bottom of the chromosphere and consequently the vertical plasma velocity turns out to be the ramp function of time $R(t)=t H(t)$ whereas the source term $S(x,y,t)$ for the density follows the delta function $\delta(t)$ form. The upward acceleration ${\bf a}=a(x,y)\hat{z}$ produced in this density structure is greater than the downward constant solar acceleration $-{\bf g}_\odot$ in the chromosphere.  In the transition region (TR), the temperature gradients are steeper; therefore, the upward acceleration increases in magnitude $g_\odot << a$ and the density hump spends lesser time there. This is stage-II of its life cycle. In stage-III, the density structure enters into corona where the gradients of temperatures vanish and the structure decelerates to zero velocity under the action of the solar gravitational force. 

\end{abstract}



\section{\label{sec:level1}Introduction}
Material ejection in the vertical direction from the surfaces of astronomical objects is a very common phenomenon. Small-scale plasma jets, called spicules, were first observed long ago \cite{secchi1877soleil} in the solar atmosphere. The physical mechanism for the generation of large-scale classical plasma jets has similarities with the small-scale outflows of plasma observed in different regions of the solar atmosphere \cite{shibata1986magnetodynamic, bellan2018experiments}. 
Vertical plasma jets emerging in the perpendicular direction to the solar disk can always be observed, which have lengths of the order of thousands of kilometers, diameters of the order of hundreds of kilometers and life time of the order of (5-10) minutes.  These density structures move in the upward direction with speeds of tens of kilometers per second \cite{priest1982solar}.
About half a million \cite{lang2001cambridge} or at least about hundred thousand $100,000$ \cite{de2004solar} spicules can always be seen at the solar disk with  variations in size. So far, there has been no consensus on the fundamental mechanism for the generation of astrophysical jets at different spatial scales. Several physical mechanisms can be involved in producing these jet-like flows in different astronomical environments.  

Sun is the closest astronomical object to our home, Earth, and a great variety of small-scale localized jet-like plasma flows and mass ejections have been detected in the three regions of its atmosphere; the surface, the chromosphere and the corona \cite{priest1982solar, priest2014magnetohydrodynamics, woo1996kilometre, de2007tale, aschwanden2017width, klimchuk2006solving}. It includes spicules, coronal loops, coronal mass ejections (CMEs), and the much smaller cylindrical plasma structures; the threads and strands observed within the coronal loops \cite{goddard2017statistical}. 

During the past several years, a lot of interest has been invoked in the study of solar spicules, particularly because of the advancement in observational techniques which led \cite{de2007tale} to divide the spicules into two categories; one has been named as Type-I and the other Type-II. Type-I spicules exhibit slower velocities $(10-40) km s^{-1}$ and longer life times $(3-10)$ minutes, while Type-II spicules have larger velocities $(80-300) km s^{-1}$ and shorter life times $(1-3)$ minutes \cite{de2007tale}. In addition to upward plasma motions, downward flows have also been observed in the transition region (TR) and lower chromosphere. High-speed downward flows with velocities $(60-200) km s^{-1}$ occur near the active regions as well as in the quiet Sun (Bose et al. 2021). Thermal energy flows from the chromosphere to the corona, but only a small fraction of energy escapes to the corona \cite{carlsson2019new} and the rest remains trapped within the chromosphere. After their life time, spicules either fade out or fall back into the chromosphere \cite{bose2021spicules, scalisi2021propagation}.

In order to gain a deeper understanding of the morphology and characteristics of solar jets, a few physical mechanisms have been proposed. These mechanisms include the shock waves generated by pressure/velocity pulses \cite{hollweg1982origin, shibata1982numerical, kuzma2017two}, shock waves driven by nonlinear Alfven waves in magnetic flux tubes \cite{matsumoto2010nonlinear} as well as the creation of jets by the magnetic reconnection  \cite{yokoyama1995magnetic, gonzalez2017jet}. The reflection of Alfven waves has also been considered as one of the possible mechanisms for the creation of spicules \cite{scalisi2021reflection}.

Numerical simulations can consider a multitude of physical effects by including the terms representing radiation, conduction, resistivity, and pressure in the set of plasma equations to investigate the evolution of jet-like flows. Series of numerical experiments have been performed using single fluid magnetohydrodynamics (MHD). The role of several parameters in determining the dynamics of spicules has been studied \cite{martinez2017generation, martinez2020ion}. The 2D simulations provide a clue to the existence of pressure force within the spicule structure, and the plasma within the boundaries is found to be in the state of non-equilibrium. The non-uniform internal pressure force and magnetic field tension counterbalance each other and cause cross-sectional deformation. Regions of high density and pressure have been given the names of knots by some authors \cite{dover2021magnetohydrodynamic}. A new phenomenon of nanojets has also been observed recently \cite{antolin2021reconnection}.

The creation and dynamical evolution of the spicules have also been investigated through numerical simulations by launching the vertical velocity from the upper chromosphere using MHD equations \cite{kuzma2017two}. The set of equations used in this simulation reduces to MHD if neutrals are ignored. The authors have considered several physical effects, such as gravitational field, ionization, recombination, collisions, and pressure gradient forces. The current-free plasma is considered and in equilibrium the pressure gradient balances the gravity. Then a time-dependent perturbation is introduced at the bottom of the plasma boundary to investigate the temporal evolution of densities and pressures. The results indicate a rapid increase in the plasma density with height which is not in complete agreement with the observations. However, several other features indicated by the simulations match with the observations. The MHD simulations \cite{dover2021magnetohydrodynamic, kuzma2017two} predict the role of pressure gradients in the formation and evolution of spicules, and they do not rely only on the magnetic reconnection mechanism. Numerical simulation has also been performed for the development of spicules with velocity pulse generated in the chromosphere using ideal and non-ideal MHD models including pressure force \cite{kuzma2017numerical}.

Long ago \cite{biermann1950ursprung}, it was proposed that the electron baroclinic vector $(\nabla n_e \times \nabla T_e)$ generates the seed magnetic fields in stars. 
Ions were assumed to be static $(m_i \rightarrow \infty)$ and electrons were considered to be inertialess $(m_e \rightarrow 0)$ in this theoretical model. The spatial variations of density and temperature were assumed to be constant in time and the electron Lorentz force term was also ignored.
The large magnetic fields observed in classical laser plasma experiments of the order of megagauss ($10^{6}$) $ G$ \cite{brueckner1974laser} were explained on the basis of the Biermann battery mechanism. Keeping in view the intermediate temporal and spatial scales for the case $m_i \rightarrow \infty$ and $m_e \rightarrow 0$, a theoretical model named electron magnetohydrodynamics (EMHD) was presented \cite{kingsep1990reviews, bol1991handbook}. This mechanism was also used to estimate the seed fields in galaxies of the order of ($3 \times 10^{-17}$ $G$) \cite{lazarian1992diffusion, widrow2002origin}. 
It may be mentioned here that the baroclinic flow, baroclinic waves and instabilities are well-known in atmospheric physics. The equations used in different models are solved under appropriate boundary conditions applicable in the atmosphere \cite{de2020barotropic, baran2022numerical, agaoglou2024building}.

It was shown that the assumption of stationary ions in Biermann mechanism and in EMHD model was not convincing for the study of the problem of magnetic field generation  \cite{saleem1996theory, saleem1999electron}. Therefore, the Biermann mechanism was modified by including ion dynamics and plasma flows to generate the coupled seed magnetic field ${\bf B}$ and plasma vorticity $(\nabla \times {\bf v}_i)$. The 2D exact solutions of two fluid plasma equations including flows of electrons and ions were found assuming spatial variation of density and temperatures to be constant with respect to time in Cartesian geometry \cite{saleem2007theory, saleem2010nonequilibrium}. In these investigations, it was pointed out that the plasma vorticity is coupled with the magnetic field and both physical quantities are created simultaneously by the baroclinic vectors of electrons and ions. Hence the mechanism of magnetic field generation should not be considered independent of vorticity (or flow) generation. 
Later on \cite{saleem2021exact}, the above theoretical model was extended to explore the possibility of generation of jet-like flows by thermal forces of electrons and ions using cylindrical coordinates. It was shown that plasma flow in the perpendicular direction to the surface is generated if the density attains locally the spatial profile like the Bessel function of order one in the radial direction and the temperatures vary linearly along the axial direction In this theoretical model, the jet-like flows are produced in both directions; upward and downward because the Bessel function is an oscillatory function. 

Recently, the 3D exact solutions of two fluid plasma, MHD and neutral fluid equations have been presented in Cartesian geometry \cite{saleem20223d} assuming the gradients of density and temperatures to be constant with respect to time. Following the 2D cylindrical solution of plasma equations for the creation of jet-like flows mentioned above, a model for the creation of plasma jets in Cartesian coordinates has also been presented. In that work, it has been shown that plasma flow in the upward direction is produced when the density has a spatial dependence on the surface coordinates $n=n(x,y)$ in a local region and the electron and ion temperatures have positive gradients along the vertical direction, that is, $\nabla T_j= \mid \frac{dT_j}{dz}\mid \hat{ z}$ where $j=(e,i)$. This exact analytical solution of plasma equations has been applied to explain the generation of spicules by plasma baroclinic vectors $\nabla n \times \nabla T_j$. 
The creation of the plasma slab at the bottom of the spicule can be divided into four quadrants in the xy plane. The density is assumed to be maximum at the center. Considering only one out of four quadrants, it has been shown that a slab of a small height at the bottom of the structure will be lifted upward with the acceleration greater than the solar gravitational acceleration acting downward. When the slab enters into upper regions, either the density gradient or temperature gradient vanishes and its speed becomes constant in agreement with the observations. The vorticity and magnetic field turn out to be linearly growing with time by the time-independent baroclinic vectors of electrons and ions. But this model cannot consider the time evolution of the plasma jet at the initial stage.

Physically a density hump is created in space during a finite time 't' however small it may be.
If constant temperature gradients are given as is approximately true in different regions of the solar atmosphere \cite{slemzin2014spectroscopic, priest2014magnetohydrodynamics}, then wherever a density hump or dip is created locally as a function of $(x,y)$, it must be time-dependent at least at initial evolution stage of the ordered plasma structure; the plasma jet. A more realistic solution of plasma equations must have a time-dependent density $n=n(x,y,t)$ to produce upward flows locally in the presence of constant temperature gradients along $z$ axis.

Our task is two fold; first we find the exact analytical solution of two fluid plasma equations with time-dependent density $n=n(x,y,t)$ and constant gradients of temperatures along the positive z-axis $T_j=T_j(z)$. This solution can also be used to investigate the generation of large-scale plasma jets emerging from YSOs and other astronomical objects in the classical limit. Second, we apply the above mentioned 2D solution of two fluid plasma equations to explain the evolution of spicules and their life cycle. The birth and death of spicules are explained by considering three stages in three different regions; chromosphere, transition region (TR), and lower corona. The 2D time-dependent density structure presented in this work will be used in the evolution stage of the spicule in the chromosphere. Then the 2D solution \cite{saleem20223d} obtained for time-independent density will be employed to elaborate the dynamics of the spicule in TR and corona using basic classical kinematics.

\section{Theoretical Model for Plasma Jet Formation}

Let us consider two fluid ideal plasma equations of momentum  conservation for electrons and ions in the presence of constant gravitational acceleration ${\bf g}$ which are written, respectively, as \cite{chen1984introduction},
\begin{equation}
	m_e n_e (\partial_t + {\bf v}_e \cdot \nabla) {\bf v}_e = -en_e ({\bf E}+\frac{1}{c} {\bf v}_e \times {\bf B}) 
- \nabla p_e + m_e n_e {\bf g} 
\end{equation}
and
\begin{equation}
	m_i n_i (\partial_t + {\bf v}_i \cdot \nabla) {\bf v}_i = e n_i ({\bf E}+\frac{1}{c} {\bf v}_i \times {\bf B}) - \nabla p_i + m_i n_i {\bf g}  
\end{equation}
Since spicules have an ambient magnetic field created by solar plasma dynamics, therefore we add a unidirectional constant magnetic field ${\bf B}_0= B_0 \hat{z}$ to the weak magnetic field generated by the baroclinic vectors ${\bf B}_(x,y,t) = \tilde{\bf B}$ and write the total field within the structures ${\bf B}_{T}$ as,
\begin{equation}
	{\bf B}_{T}={\bf B}(x,y,t)+{\bf B}_0 
\end{equation}
Both electrons and ions are assumed to obey the ideal gas law i.e.$p_j = n_j T_j$ where subscript $j=(e,i)$ denotes electrons and ions. It will be shown that when a density hump is created within a local region of the chromosphere in xy-plane in the presence of constant gradients of the temperatures along ${\bf B}_0$, it will be accelerated upward by the force of baroclinic vectors $\nabla n \times \nabla T_j$. The quasi-neutral plasma $n_e \simeq n_i=n(x,y,t)$ is in a state of non-equilibrium $T_e \neq T_i$. The continuity equations require a source term $S(x,y,t)$ and they can be written for jth species in the following form,
\begin{equation}
	\partial_t n + \nabla \cdot (n {\bf v}_j) = S(x,y,t)
\end{equation}
Since the current density is ${\bf j}= e n ({\bf v}_i-{\bf v}_e)$, therefore Amperes' law gives electron velocity in terms of ions velocity,
\begin{equation}
	{\bf v}_e = {\bf v}_i - \frac{c}{4 \pi e}(\frac{\nabla \times {\bf B}}{n})
\end{equation}
The equations of motion of electrons and ions for longitudinally uniform flow $\nabla \cdot {\bf v}_j =0$ take the form of following two coupled equations \cite{saleem2010nonequilibrium}, respectively,
\begin{equation}\label{eom1}
	\partial_t {\bf B}=\nabla \times ({\bf v}_i \times
	{\bf B}) +\nabla \times ({\bf v}_i \times
	{\bf B}_0)-(\frac{c}{4\pi n e}) \{\nabla \times [(\nabla \times {\bf
		B})\times {\bf B}]\}
\end{equation}
$$
-(\frac{c}{4\pi n e}) \{\nabla \times [(\nabla \times {\bf
	B})\times {\bf B}_0]\}+\frac{c}{4 \pi n e} \{\nabla \psi\times [(\nabla \times {\bf
	B})\times {\bf B}]\}
$$
$$
 +\frac{c}{4 \pi n e} \{\nabla \psi\times [(\nabla \times {\bf
 	B})\times {\bf B}_0]\} -\frac{c}{e} (\nabla\psi \times \nabla T_e)
$$
and 
\begin{equation}\label{eom2}
	a \partial_t {\bf B}+\partial_t(\nabla \times{\bf v}_i)=\nabla
	\times [{\bf v}_i\times(\nabla \times{\bf v}_i)]+a\nabla\times({\bf
		v}_i\times{\bf B})
\end{equation}
$$
+a\nabla\times({\bf
	v}_i\times{\bf B}_0)+\frac{1}{m_i}(\nabla \psi \times \nabla
T_i)
$$
where $a=\frac{e}{m_i c}$, $\psi=\ln \bar{n}$, $\bar{n}=\frac{n}{N_0}$, and $N_0$ is
arbitrary constant density. 
If following two conditions also hold,
\begin{equation}
	\nabla \psi \cdot {\bf v}_i=0 
\end{equation}
and
\begin{equation}
	\nabla \psi \cdot (\nabla \times {\bf B})=0 
\end{equation}
then all nonlinear and complicated terms of Eqs. (\ref{eom1}) and (\ref{eom2}) vanish and they reduce to two simpler equations 
\cite{saleem2010nonequilibrium,saleem2021exact,saleem20223d}, respectively, given as,
\begin{equation}\label{b}
	\partial_t {\bf B}=-\frac{c}{e} (\nabla \psi \times \nabla T_e) 
\end{equation}
and
\begin{equation}\label{vb}
	\frac{e}{m_i c} \partial_t {\bf B}+\partial_t(\nabla \times {\bf
		v}_i)=\frac{1}{m_i}(\nabla \psi \times \nabla T_i)
\end{equation}
We assume $\nabla \psi$ is not parallel to $\nabla T_j$. Then (10) and (11) give an expression for the generation of ions vorticity by baroclinic vectors,
\begin{equation}\label{v}
	\partial_t (\nabla \times {\bf v}_i)= \frac{1}{m_i} \nabla \psi \times (\nabla T_{e} + \nabla T_{i}) 
\end{equation}
Equation (\ref{v})  implies that the thermal force of plasma is the source for the generation of ions vorticity which is coupled with the electrons vorticity through Eq. (5).

\subsection{Generation of Plasma Jet}
If $\psi = \psi(x,y,t)$ and $T_j= T_j(z)$, then the right hand side of Eq. (12) can be expressed as,
\begin{equation}
\frac{1}{m_i} \nabla \psi \times (\nabla T_{e} + \nabla T_{i})=a_0 (\partial_y \psi, -\partial_x \psi, 0)   
\end{equation}
where $a_0= (\frac{T_{e}^{\prime}+T_{i}^{\prime}}{m_i})$ and $T_{j}^{\prime}=\frac{d T_j}{d z}$. We have $\nabla \times {\bf v}_i=[(\partial_y v_{iz}-\partial_z v_{iy}), (\partial_z v_{ix}-\partial_x v_{iz}),(\partial_x v_{iy}-\partial_y v_{ix})]$ and relation (13) demands that $\partial_x v_{iy}-\partial_y v_{ix}=0$, in general, which is possible if $v_{ix}=v_{iy}=0$. In this case, Eq. (12) becomes,
\begin{equation}
\partial_t (\partial_y v_{iz}, - \partial_x v_{iz}, 0)=a_0 (\partial_y \psi, -\partial_x \psi, 0)
\end{equation} 
The above relation shows that the ions velocity has non-zero component only in the direction perpendicular to the plasma surface that is along z-axis.
The temperatures in solar atmosphere vary along z-axis almost linearly in different layers \cite{slemzin2014spectroscopic, priest2014magnetohydrodynamics}, therefore we assume,
\begin{equation}
T_j = T_{j}^{\prime} z + T_{0j}
\end{equation}
where $T_{j}^{\prime}$ and $T_{0j}$ are constants. Density is assumed to be the spatial function of the coordinates $(x,y)$ and time "t", therefore $\psi=\psi(x,y,t)$. The ions velocity is determined by the Eqs. (12) and (13) in the following form,
\begin{equation}
{\bf v}_i= (0,0,v_{iz})=v_{i} (x,y,t) \hat{z}
\end{equation}
Equations (10) and (11) imply that the 
baroclinic vectors of ions and electrons are parallel to each other and Eq. (14) indicates that the spatial profile of $v_{i}$ is similar to $\psi$, therefore, ${\bf B}$ is also parallel to the ion vorticity $\nabla \times {\bf v}_i$. Hence, we assume,
\begin{equation}
{\bf B}= b_0 (\nabla \times {\bf v}_i)
\end{equation}
where $b_0$ is constant. The form of $v_{i}$ given by Eq. (16) yields,
\begin{equation}
\nabla \times [{\bf v}_i\times(\nabla \times{\bf v}_i)]=0
\end{equation}
Equations (17) and (18) imply,
\begin{equation}
\nabla\times({\bf v}_i\times{\bf B}) =0
\end{equation}
Equation (17) yields,
\begin{equation}
\nabla \times {\bf B} = -b_0 \nabla^2 {\bf v}_i
\end{equation}
If 
\begin{equation}
\nabla^2 {\bf v}_i = \eta {\bf v}_i
\end{equation}
where $\eta$ is constant, then both generated fields ${\bf B}$ and ${\bf v}_i$ turn out to be curls of each other. Equations (14) and (21) demand,
\begin{equation}
	\nabla^2 \psi = \eta \psi
\end{equation}

\section{Analytical Solutions}

The general solutions of Eqs. (10), (11) and (12) depend upon the form of $\psi$ in the presence of spatial gradients of temperatures along the z direction defined by Eq. (15). We look for an exact analytical solution of two fluid equations with a time-dependent density function $\psi=\psi(x,y,t)$. 

\subsection{The $\psi$ as a Step Function in Time}
Let us consider the physical phenomenon of the the creation of a density hump in chromosphere at initial time $t=0$ with two dimensional spatial form $F(x,y)$,
\begin{equation}
F(x,y) = (A_1 e^{-(\mu x + \nu y)}+ A_2 e^{(\mu x - \nu y)} )
\end{equation}
where $A_1$, $\mu$, $\nu$, $A_2$ are constants. 
Let the source term in Eqs. (4) be,
\begin{equation}
S(x,y,t)= n F(x,y) \delta(t)
\end{equation}
Then continuity equations (4) reduce to,
\begin{equation}
\partial_t \psi = F(x,y) \delta(t)
\end{equation}
The above relation defines $\psi$ as follows,
\begin{equation}
\psi(x,y,t) =  F(x,y) H(t)
\end{equation}
where $H(t)$ is the step function and hence $H(t)=1$ for $0\leq t$ which gives, 
\begin{equation}
\psi(x,y,0) = F(x,y)
\end{equation}
The Eq. (23) implies,
\begin{equation}
\nabla^2 F = \eta F
\end{equation}
where $\eta=(\mu^2 + \nu^2)$ and it is in agreement with Eq. (22). Equation (14) can be written in terms of $F$ and $H(t)$ using the definition of function $\psi$,
\begin{equation}
\partial_t (\partial_y v_{i}, - \partial_x v_{i}, 0)=a_0 (\partial_y \psi, -\partial_x \psi, 0)=a_0 (\partial_y F, -\partial_x F, 0) H(t)
\end{equation}
The above equation implies that $v_i$ should have the spatial profile similar to $F$.
\begin{equation}
\partial_t v_{i}(x,y,t)  =a_0 F(x,y) H(t)=a(x,y) H(t)
\end{equation}
Integration of above equation with the initial condition $v_{i}=0$ at $t=0$, yields the ions vertical flow as,
\begin{equation}
v_{i}(x,y,t)  =a_0 F(x,y) R(t)=a(x,y) t
\end{equation}
where $R(t)=t H(t)$ is the ramp function which obeys the following relation,
\begin{equation}
\frac{d}{dt} R(t) = H(t)
\end{equation}
Equations (16) and (31) determine the acceleration in  vertical direction,
\begin{equation}
{\bf a}(x,y) = a(x,y) \hat{z}=a_0 F(x,y) \hat{z}
\end{equation}
which is also a function of $(x,y)$ coordinates. The density profile will move upward if the condition $g_{\odot}<a_0 F(x,y)$ holds.
Using Eq. (10) along with Eq. (26), the magnetic field generated by baroclinic vectors can be expressed in terms of functions $F$ and $H(t)$ as follows,
\begin{equation}
\partial_t {\bf B}(x,y,t)= - \frac{c T_{e}^{\prime}}{e}(\partial_y \psi, - \partial_x \psi, 0)= - \frac{c T_{e}^{\prime}}{e}(\partial_y F, - \partial_x F, 0) H(t)
\end{equation}
Integration of above equation with the condition $B(x,y,0)=0$ at $t=0$ gives the magnetic field created by baroclinic vectors as a ramp function of time,
\begin{equation}
{\bf B}_(x,y,t)= -\frac{c T_{e}^{\prime}}{e} (\partial_y F, - \partial_x F,0)R(t)
\end{equation}

\bigskip

\begin{figure}
    \centering    \includegraphics[width=10cm, height=8cm]{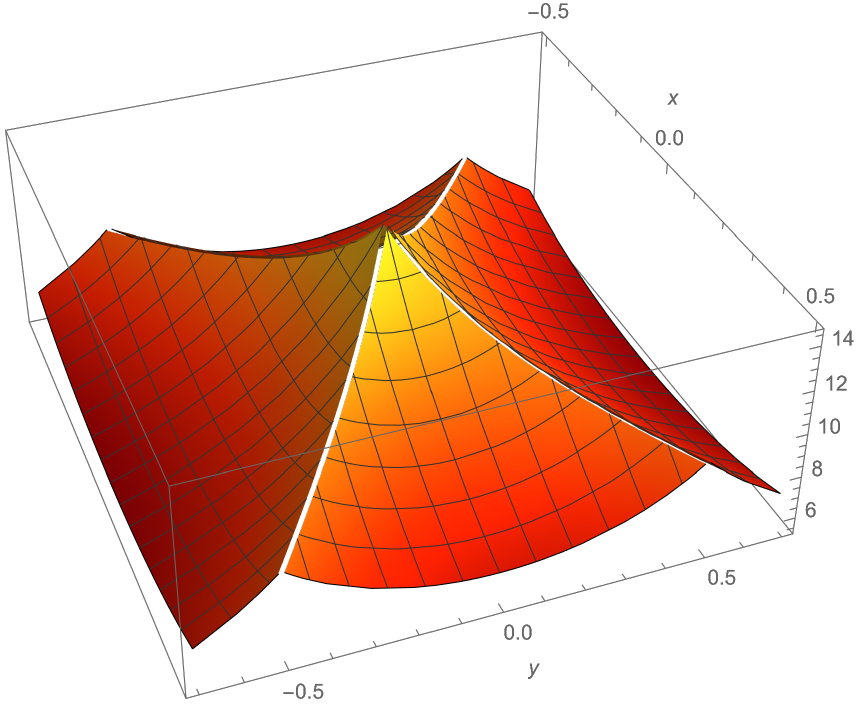}
    \caption{The function used to plot the above figure is $H(x+y) e^{\psi_0 e^{(-\mu x -\nu y)}} + H(-x+y) e^{\psi_0 e^{(\mu x -\nu y)}} + H(x-y) e^{\psi_0 e^{(-\mu x +\nu y)}}+ H(-x-y) e^{\psi_0 e^{(\mu x +\nu y)}}$ and here $H(x,y)$ is the Heaviside function on 2D space. The above form of the function allows us to plot the solution in all four quadrants. It represents the density profile in xy-plane in the simpler case corresponding to $A_2=0$ in Eq.(23).} 
    \label{fig1}
\end{figure}

\subsection{The $\psi$ as an Exponential Function in Time}

It may be mentioned here that we can also get a mathematically consistent solution by considering, 
\begin{equation}
\psi(x,y,t) = F(x,y) e^{\gamma t}
\end{equation}
where $\gamma$ is a constant. 

In this case, we can write Eq. (10) as,
\begin{equation}
\partial_t {\bf B}(x,y,t)= - \frac{c T_{e}^{\prime}}{e}(\partial_y \psi, - \partial_x \psi, 0)= - \frac{c T_{e}^{\prime}}{e}(\partial_y F, - \partial_x F, 0) e^{\gamma t}
\end{equation}
Integrating above equation with the initial condition $B(x,y,0)=0$ at $t=0$, we obtain,
\begin{equation}
	{\bf B}(x,y,t)=
	- \frac{c T_{e}^{\prime}}{e \gamma}  (\partial_y F, -\partial_x F, 0) (e^{\gamma t} -1)
\end{equation}
The ions vorticity equation (12) becomes,
\begin{equation}
\partial_t (\nabla \times {\bf v}_i)=\partial_t (\partial_y v_{i}, - \partial_x v_{i}, 0) =a_0 (\partial_y F, - \partial_x F, 0) e^{\gamma t}
\end{equation}
Integrating above equation with the initial condition $v_{iz}(x,y,0)=0$, we obtain the jet-like flow  ${\bf v}_i$ to be generated in the following form,
\begin{equation}
	{\bf v}_i= \frac{a_0 }{\gamma} F(x,y) (e^{\gamma t} -1) \hat{z}
\end{equation}
In this case, the density source can be defined as,
\begin{equation}
	S(x,y,t)= =\gamma n(x,y,t) (e^{\gamma t} \psi_0 F(x,y))=\gamma n \psi
\end{equation}
The continuity equations reduce to,
\begin{equation}
	\partial_t \psi = \gamma \psi
\end{equation}
At $t=0$, Eq.(36) becomes Eq. (27).
Since density becomes double exponential function of time, therefore the parameter $\gamma$ must be chosen very carefully. Note that in this case, the density $n(x,y,t)$ increases very rapidly with time and this solution can be valid only over an extremely short duration of time. 
Thus, the choice of step function form of $\psi$ during evolution time seems to be physical and we will use it to investigate the birth and death of spicules in the next section.\\

\section{Creation and Life Cycle of Spicules}

Let us suppose that a density hump is created in the form of a rectangle within the chromosphere with area $\sigma=2x_m \times 2y_m$ where $-x_m\leq x \leq x_m$ and $-y_m\leq x \leq y_m$. Keeping in view the observed order of the spatial dimensions of the spicules \cite{priest1982solar, priest2014magnetohydrodynamics}, we choose $x_m=(200) km$ and $y_m=(300) km$ .
The density function $F$ is maximum at the point $(0,0)$ and decreases smoothly within the rectangular area away from the centre for $A_2<<A_1$. 

Since, we define density as a function of two coordinates $(x,y)$, the density profile does not follow the exact exponential form contrary to one dimensional (1-D) case used for the study of electrostatic drift waves in magnetized plasmas.
It may be mentioned here that the equilibrium density gradient $\nabla n_0$ (where $n_0$ denotes equilibrium density) in magnetized plasmas is assumed to be one-dimensional (1-D), in general, to study the linear and nonlinear waves. For example, the electrostatic drift wave is an important low frequency wave which gives rise to instabilities, energy transport and nonlinear structures in plasmas. In most of the research work on that topic, the equilibrium density variation is assumed to be of the exponential form i.e. $n(x)= n_0 e^{-\frac{x}{L_n}}$ where $\mid \frac{1}{n_0} \frac{d n_0} {dx}\mid= \kappa_n=\frac{1}{L_n}=$ constant. The equilibrium density decays exponentially away from the point $x=0$ where $L_n$ is the density gradient scale length \cite{weiland1999collective}.
These waves were first studied theoretically
long time ago \cite{kadomtsev1963anomalous} and the same year they were observed in laboratory \cite{d1963low}. Later, the
electrostatic turbulence in magnetized plasma was explained by the nonlinear drift waves \cite{hasegawa1978pseudo}. If ambient magnetic field is along the z-direction ${\bf B}_0=B_0 \hat{z}$, then this longitudinal wave propagates predominantly along positive y-axis. However, in the case of spicules, the density must have at least two-dimensional (2-D) variation. 
	
The spicules originate from different regions of active and quiet Sun and have been divided into two main types. The type-I spicules have typical life times of the order of $3$ to $10$ minutes, diameters of $120$ to $700$ $km$, heights of 
$3$ to $5$ $Mm$, temperatures $10,000$ to $15,000$ $K$ electron densities $10^{17}$ $m^{-3}$ and reach the speeds $10$ to $50$ $km s^{-1}$. Type-II spicules reach speeds of $30$ to $150$ $km s^{-1}$ and have longer heights compared to Type-I and are shorter in active regions ($2$ to $5$ $Mm$) and taller in coronal holes $5$ to $10$ $Mm$ \cite{priest2014magnetohydrodynamics, beckers1972solar, de2007tale}. On the other hand, the macrospicules are bigger structures with heights $4$ to $40$ $Mm$, diameters $4$ to $11$ $Mm$ and have life time of $8$ to $45$ minutes \cite{priest2014magnetohydrodynamics}.
Multitude of theoretical models have been presented for understanding the dynamics of spicules and causes of their creation \cite{hollweg1982origin, de2014interface, iijima2015effect, shimojo2020estimating}. 
There is no consensus on any single complete theory for the creation of spicules and complete comparisons between simulation results and observations have not been presented. Furthermore, the observations do not give the detailed changes of density of spicules at different altitudes. The present theoretical model also does not take into account the density dependence on z-coordinate.
We assume that the spicules are created in chromosphere, pass through transition region and rise to heights of thousands of kilometers in corona. At every point within the spicule, the upward acceleration ${\bf a}(x,y)$ and velocity ${\bf v}_{i}(x,y)$ have different magnitudes. These three stages are discussed using the exact solutions of two fluid plasma equations. 

In the solar atmosphere, the temperature first decreases from the surface value $T \simeq 6600$ $K$ to $T \simeq (4300)$ $K$ at an altitude of about $500$ $km$ and then  gradually increases from $T\simeq (4300)$ $K$ to $(2.5) \times 10^{4}$ $K$  in the upper chromosphere \cite{priest2014magnetohydrodynamics} and after passing through the thin transition region (TR), the temperature increases rapidly to $T\simeq (2-3) \times 10^{5}$ $K$ \cite{slemzin2014spectroscopic}. Above transition region lies the lower corona where the temperature reaches the value of the order of $T \simeq 10^{6}$ $K$ and its gradient becomes negligibly small. 

The plasma in chromosphere is not in thermal equilibrium and several physical phenomena such as ionization, recombination, radiation emission and collisions among charged-charged, charged-neutral and neutral-neutral particles are continuously taking place \cite{vranjes2013collisions, kuzma2019heating, martinez2017generation}. The data of the temperature of electrons $T_e$ and protons $T_i$ is not available separately within the region of chromosphere. However, the values of temperature $T$ at different altitudes are mentioned in literature assuming the plasma as a single fluid. We are considering only the hydrogen plasma ignoring the contribution of alpha particles, neutrals and collisions. Theoretical models from mid chromosphere to the corona and the solar wind predict that the proton temperature is higher than the electron temperature $T_e < T_i$ \cite{hansteen1997role}. In the light of these models, we can expect the condition $T_e < T_i$ also holds in chromosphere. Thus we choose $T_i= 2T_e$ and interestingly this relation of temperatures yields reasonably good estimates of the magnitudes of upward accelerations, flows and the times spent by the density hump in chromosphere, transition region and corona. It is important to mention that in the present theoretical model, the acceleration ${\bf a}$, upward plasma flow ${\bf v}$ and times $\tau$ (spent by the spicule in different regions) are functions of $F(x,y)$ and hence depend upon $(x,y)$ coordinates. The values of these quantities have been estimated at two points; the centre point $(0,0)$ and a point near the edge $(x_m, y_m)$ of spicule. It may be noted that we get  $\psi =0$ at the points where $n(x,y,t)=N_0$ and the phenomenon of flow generation disappears. The vertical velocities in each region depend upon the form of $F$ which is a function of $(x,y)$ coordinates, therefore the density structure has a non-uniform vertical flow. 

\subsection{Life Cycle of Spicule}
Keeping in view the above mentioned temperatures in different regions of solar atmosphere and their thicknesses, we divide the spicule generation process into three stages. The first stage is its birth in the lower chromosphere where the density hump in $(x,y)$ coordinates is created due to the plasma dynamics in the presence of constant vertical temperature gradients. The thermal force is proportional to  $\nabla \psi \times (\nabla T_{e} + \nabla T_{i})$ and it produces the ion vorticity that is associated with their vertical flow. It is assumed that the density structure formed at $t=0$ is accelerated in upward direction and spends time $\tau_{ch}$ in the chromosphere before leaving it after traveling the distance  $1500$ $km$. It then crosses the thin transition region in a short time $\tau_{tr}$ as the acceleration produced in TR is larger due to the larger temperature gradients whereas thickness of the layer is small. After gaining a larger velocity in TR, it enters the corona where $\nabla T_j \rightarrow 0$ and there is no upward force acting on the structure. The solar gravitational acceleration $(- g_{\odot})$ then slowly brings the velocity of the structure to zero after a time $\tau_{cr}$. The total life time of the spicule is $\tau_l=(\tau_{ch}+\tau_{tr}+\tau_{cr})$.

\subsection{Creation of Spicule in Chromosphere}
Let us estimate the magnitudes of the acceleration and velocity of the density hump in this area during time $t=0 \rightarrow \tau_{ch}$. Four corner points of this part of the area are $(0,0)$, $(x_m,0)$, $(0,y_m)$ and $(x_m,y_m)$ and we choose dimensionless parameters $\mu x_m=(0.3)$ and $\nu y_m=(0.4)$ along with $\frac{n(0,0,0)}{N_0}=5$. We find $\mu=(1.5)\times 10^{-6}$ $m^{-1}$ and $\nu=(1.33)\times 10^{-6}$ $m^{-1}$ for $x_m = (200 \times 10^{3})$ $ m$ and  $y_m = (300 \times 10^{3})$ $m$. The function $F(x,y)$ is flexible and the constants $A_1, A_2, \mu, \nu$ can have different values in different cases. Furthermore, the ratios of the densities at points $(0,0)$ and at $(x_m, y_m)$ can be varied and hence the present theoretical model suggests that vertical structures of different sizes can be created.
The above choice of the values of the constants  produces values of the upward acceleration at all four points $[(0,0),(x_m,0),(0,y_m),(x_m,y_m)]$ greater than the solar gravity i.e. $g_{\odot}< a(x,y)$. Therefore, the density hump starts moving upward and attains observable velocity. The magnitude of the upward acceleration $a(x,y)$ created by the baroclinic vectors depends also upon the magnitude of the temperature gradients. The formation of spicules in chromosphere under the framework of present theoretical model is based on the following considerations.\\

I. Let $n_0$ be the backgound density in chromosphere. Then $n_0 \leq N_0$ while we have chosen $\frac{n(0,0)}{N_0}=5$, therefore the density within the spicule can be even larger than five times the background density. But this ratio ensures that the whole part of the area in first quadrant is uplifted because $g_{\odot} < a(x,y)$ at all four corner points. The function $F$ defined by Eq. (23) with $A_2=0.1 A_1$, gives $F(0,0)=(1.1) A_1$ while we assume $\psi(0,0,0)= \ln 5 = \simeq (1.61)$ and comparing these two values, we obtain $A_1 \simeq (1.46)$ because $\psi(x,y,0)=F(x,y)$.  \\

II. The profile of density structure is proportional to $F(x,y)$ and it does not change with $z$ coordinate. This assumption is a limitation of the exact analytical solutions of two fluid equations.\\

III. The continuity equations have the source term with $\delta(t)$ and it forces the $\psi$ function to take the form of a step function $H(t)$. 
The density variation is not unidirectional and it does not have exact exponential form as it is assumed for the investigation of electrostatic drift waves.  The upward acceleration ${\bf a}$, ion velocity ${\bf v}_i$, and times $\tau$ spent by the density structure in different regions are functions of $(x,y)$ coordinates.  \\

IV. The magnetic field  generated by thermal forces turns out to be much smaller than the constant ambient magnetic field within the spicule i.e. $\mid {\bf B}(x,y,t) \mid << B_0$.\\

V. Density structures are created one after the other in the chromosphere until the final velocity of the first such structure in the corona becomes zero. At this stage, the formation of the density lump at the bottom of the spicule stops.\\

The plasma temperatures $T$ at about $500$ $km$ and $2000$ $km$ altitudes above the solar surface are given, respectively, as $T_{e}(z_1)=4.3 \times 10^3\;$K and $T_{e}(z_2)= (2.5) \times 10^4\;$ $K$ \cite{priest2014magnetohydrodynamics} and we assume the plasma temperature to be equal to the electron temperature i.e. $T=T_{e}$. Accordingly, the electron temperature gradient in chromosphere can be estimated approximately as follows,
\begin{equation}
	[T_{e}^{\prime}]_{ch} \simeq\frac{T_{e}(z_2)-T_{e}(z_1)}{(z_2-z_1)}=(19.04) \times 10^{-26}\;\hbox{J/m}. 
\end{equation}
where $(z_2-z_1)=(1.5) \times 10^{3}$ $km$.
Assuming $T_{i} \simeq 2 \;T_{e}$ for hydrogen plasma $m_i=1.67 \times 10^{-27}$ $kg$, we obtain,
\begin{equation}
	[a_0]_{ch}=\frac{T_{i}^{\prime}+T_{0e}^{\prime}}{m_i}\simeq (342.10) m/s^2
\end{equation}
Using Eq. (33), the net upward acceleration in chromosphere becomes,
\begin{equation}
	[a(x,y)]_{ch}=(a_0)_{ch}  F(x,y)-g_{\odot}
\end{equation}
where $g_{\odot}=(274)$ $m/s^2$ \cite{priest1982solar} is magnitude of the gravitational acceleration in the downward direction at the solar surface. 
At $t=\tau_{ch}(x,y)$, we get the vertical velocity of the density hump in chromosphere before entering the TR,
\begin{equation}
	[v_{i}(x,y)]_{ch} = [a(x,y)]_{ch} \tau_{ch} 
\end{equation} 
where the initial velocity at $t=0$ has been assumed to be zero.
The vertical distance to be travelled by the density is $S_{ch}=(z_2-z_1)=1500$ $km$ and hence the following relation can be used to estimate the time spent by the density hump in chromospeher,
\begin{equation}
S_{ch} = (v_i) \tau_{ch} + \frac{1}{2} a(x,y) \tau_{ch}^2
\end{equation}
We estimate the magnitudes of velocities and the spent times of the structure only at two points $(0,0)$ and $(x_m,y_m)$. From Eq. (45), we find $[a(0,0)]_{ch}= 275.42$ $m/s^2$ and $[a(x_m,y_m)]_{ch}= 19.22$ $m/s^2$.
The spent time $[\tau(x,y)]_{ch}$ as well as initial and final velocities at points $(0,0)$ and $(x_m,y_m)$ are mentioned in Table-1.

\begin{figure}
    \hspace{3.5cm}
    \includegraphics[width=17cm, height=9cm]{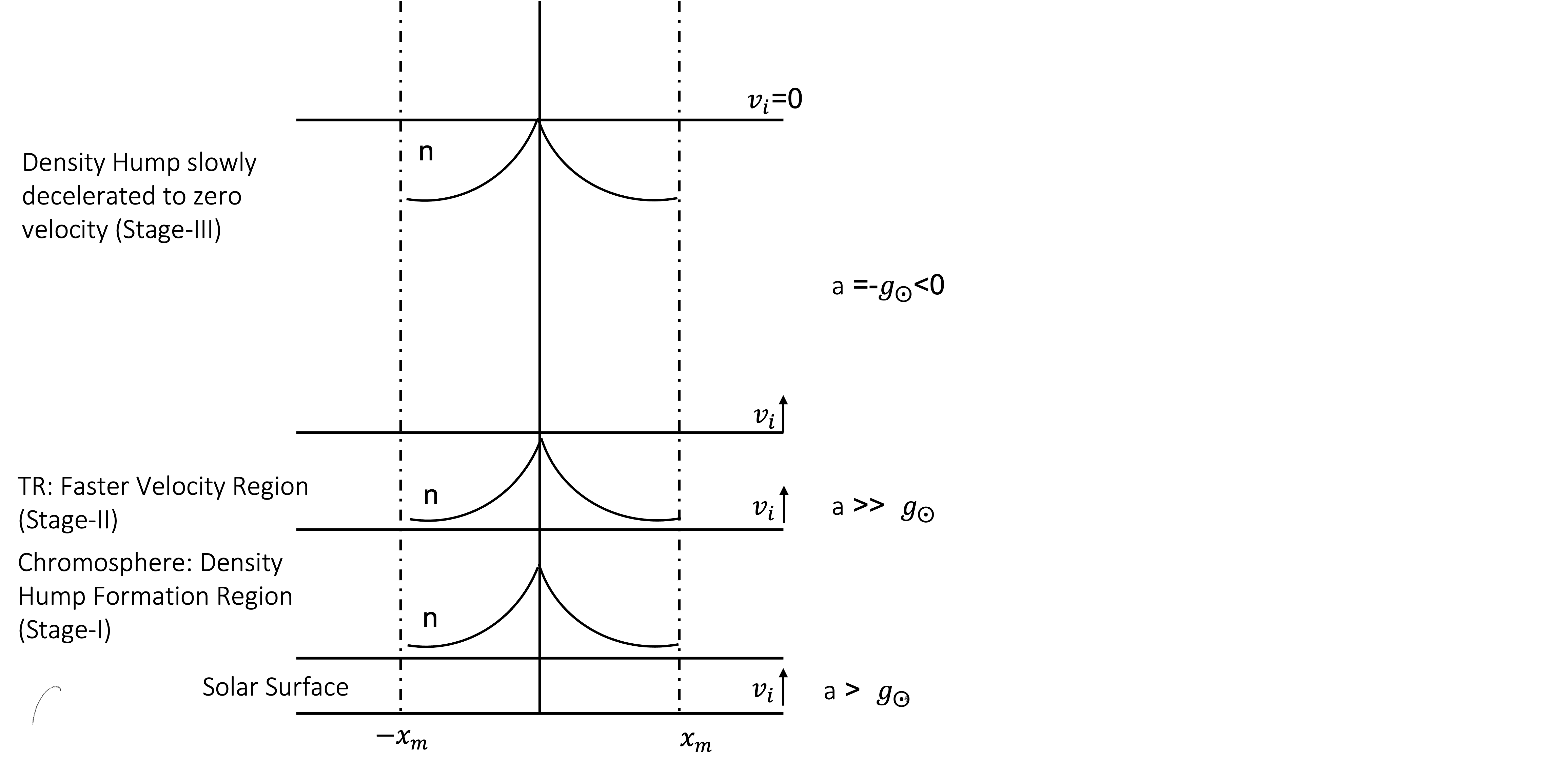}
    \caption{Schematic diagram of density hump and its velocity in three different regions of solar atmosphere. The time axis is in the upward direction.}
    \label{fig2}
\end{figure}

\subsection{Velocity Gain in the Transition Region}

The value of the temperature at top of transition region, say at point $z_3$ is given $T_{e}(z_3)=(2-3) \times 10^5\;$ $K$ (Slemzin 2014) while the thickness of this layer is about $100\;$ $km$. We take $T_e(z_3)= 3 \times 10^{5}$ $K$ and estimate,
\begin{equation}
	T_{e}^{\prime}\simeq\frac{T_{e}(z_3)-T_{e}(z_2)}{(z_3-z_2)}=(3.79) \times 10^{-23}\;\hbox{J/m}. \label{7}
\end{equation}
where $(z_3-z_2)=100$ $km$. It gives,
\begin{equation}
	(a_0)_{tr}=\frac{T_{i}^{\prime}+T_{e}^{\prime}}{m_i}=(68)  km/s^2 
\end{equation}
The net upward acceleration is,
\begin{equation}
	[a(x,y)]_{tr} = [a_0]_{tr} F(x,y) -g_{\odot} \simeq [a_0]_{tr} F(x,y)
\end{equation}
because in this region $g_{\odot}<< [a_0]_{tr} F(x,y)$.
We obtain $[a(0,0)]_{tr}\simeq (109.21) $ $km/s^2$ and $[a(x_m,y_m)]_{tr} \simeq (58.15)$ $km/s^2$.
Considering the thickness of TR to be  $(100)$ $km$, the time $\tau_{tr}$ spent in TR by these two points is estimated using the following relation,
\begin{equation}
	[a(x,y)]_{tr} \tau_{tr}^2 + 2 [v_i(x,y)]_{ch} \tau_{tr}-2 \times 10^{5}=0
\end{equation}
The velocity gained in this region by ions is estimated using following equation,
\begin{equation}
	[v_{i}(x,y)]_{tr} =[v_{i}(x,y)]_{ch} + [a(x,y)]_{tr} \tau_{tr}(x,y)
\end{equation} 
The estimated values of velocities and spent time at two points $(0,0)$ and $(x_m,y_m)$ are mentioned in Table-1.

\subsection{Rise and Fall of Spicule in the Corona}

In corona, the upward acceleration vanishes because $\nabla T_j \rightarrow 0$ and the density structure enters into this region with the velocity gained in TR i.e. $[v_{i}(x,y)]_{tr}$. The final ion velocity in corona ,say $[v_{i}]_{cr}$, becomes zero after the time $\tau_{cr}$ in corona which is longer than $\tau_{ch}$ and $\tau_{tr}$. Following relation is used to estimate the time spent in corona.
\begin{equation}
	[v_{i}(x,y)]_{cr}=[v_{i}(x,y)]_{tr}-g_{\odot} \tau_{cr}(x,y) =0.
\end{equation}.
In above equation $[v_{i}(x,y)]_{cr}$ is the final velocity of ions in corona.
In this region, the downward solar acceleration does not allow the spicule to move continuously in the vertical direction. In fact, its velocity decreases slowly here because $g_{\odot}$ is small and finally the velocity becomes zero. The corresponding heights $\textit{H}_{cr}(x,y)$ attained in corona can be estimated using the relation $\textit{H}_{cr}(x,y)=\frac{1}{2}\frac{[v_i(x,y)]_{cr}^2}{g_{\odot}}$ and it gives $\textit{H}_{cr}(0,0)= (21.28) \times 10^3$ $km$  and $\textit{H}_{cr}(x_m,y_m)= (10.54) \times 10^{3}$ $km$. 
 Total height of the spicules is $H_T=(1500)+(100)+H_{cr}$ $km$. Therefore, we obtain $H_T(0,0) \simeq (2.28) \times 10^{7}$ $m$ and  $H_T(x_m,y_m) \simeq (1.21) \times 10^{7}$ $km$ Similarly, the life time can be estimated $\tau_l(0,0)=[\tau]_{ch}(0,0)+\tau]_{tr}(0,0)+\tau]_{cr}(0,0)\simeq (8.32)$ minutes and $\tau_l(x_m,y_m)\simeq (11.25)$ minutes. 

\begin{figure}
    \centering
    \includegraphics[width=12cm, height=7cm]{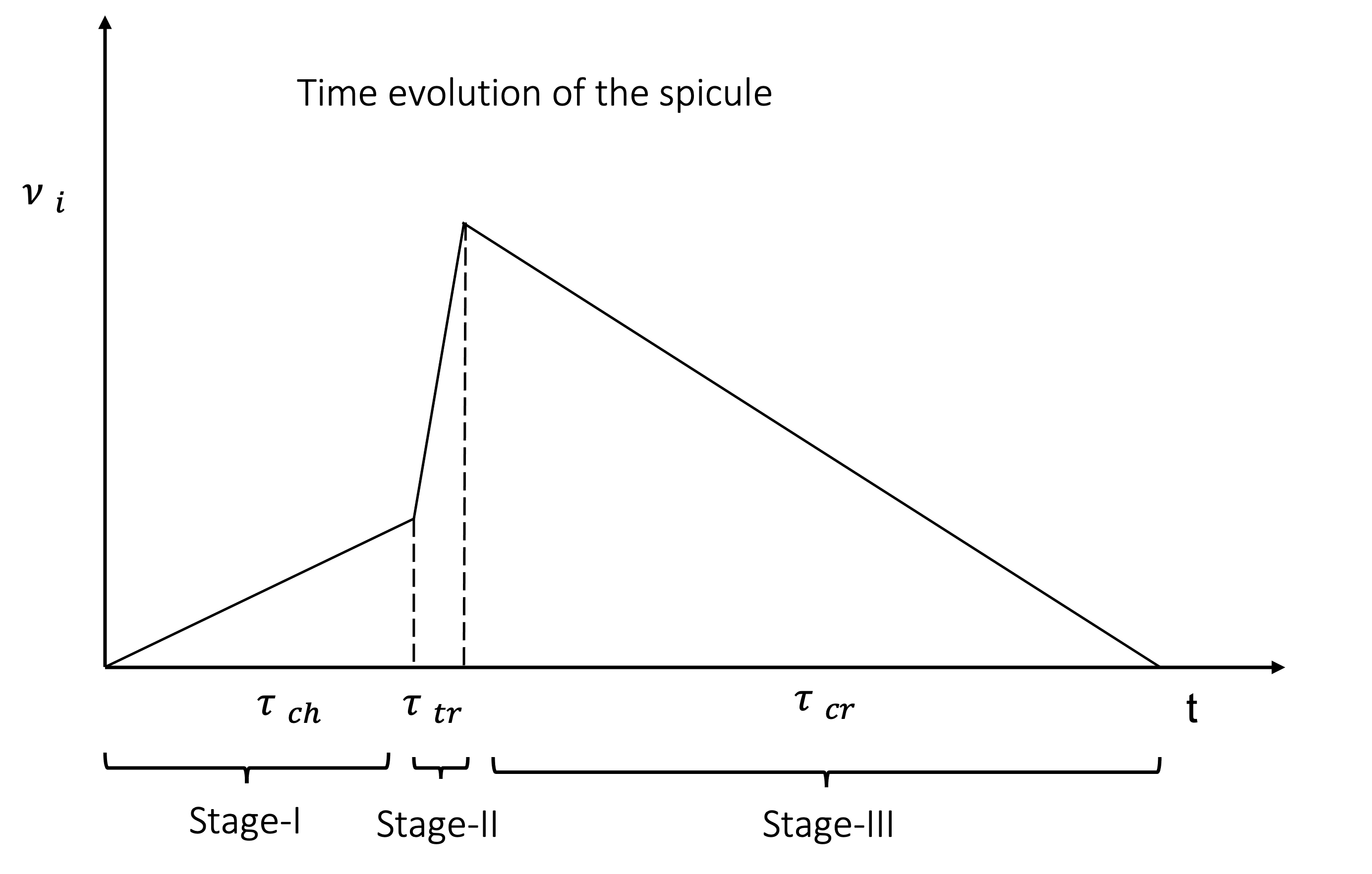}
    \caption{Schematic diagram of how the upward velocity of density hump increases slowly in stage-I, gets larger speed in stage-II and finally decreases to zero in corona after time $\tau_{cr}$. The $ \tau_{ch}$ is the time for which the spicule evolves in the chromosphere and $\tau_{tr}$ is the time the spicule spends in the transition region.   }
    \label{fig3}
\end{figure}

\subsection{Magnetic Field Generation in Spicule}
Magnitude of the generated magnetic field by electron baroclinic term can be estimated by using the electron equation. Let us evaluate it at two points chosen above in the chromosphere during $t=\tau_{ch}$. Equation (35) can be expressed in the following form,
\begin{equation}
{\bf B}(x,y,t)=(\frac{c T_e^{\prime}}{e} \nu) t [F(x,y), \frac{\mu}{\nu} A_1(- e^{-(\mu x+\nu y)}+(.1)(e^{(\mu x+ \nu y)})),0]
\end{equation}
The generated magnetic field is estimated at two points in the chromosphere as follows,
	\begin{equation}
		{\bf B}(0,0,\tau_{ch}) \simeq (1.64) \times 10^{-6} (1.61,-1.47,0) G
	\end{equation}
	and 
	\begin{equation}
		{\bf B}(x_m,y_m,\tau_{ch}) \simeq (6.24) \times 10^{-6}(0.84,-0.65,0) G
	\end{equation}
In transition region,
\begin{equation}
	{\bf B}(0,0,\tau_{tr}) \simeq (2.26) \times 10^{-6} (1.61,-1.47,0) G
\end{equation}
and 
\begin{equation}
	{\bf B}(x_m,y_m,\tau_{ch}) \simeq (3.71) \times 10^{-6}(0.84,-0.65,0) G
\end{equation}
The order of magnitude of the created magnetic field is  very small compared to the ambient magnetic field $B_0$ which varies in solar atmosphere from a few Gauss to a few hundred Gauss \cite{priest1982solar}. Since $\mu$ and $\nu$ are very small on astrophysical scales, including the solar atmosphere, the magnetic fields generated by the baroclinic vectors turn out to be very small.

It may be noted that if ions are assumed to be stationary, we get only electrons Eq. (\ref{b}) for the magnetic field generation. In this limit, Biermann battery mechanism is generalized by including electron density dependence on $(x,y)$ coordinates as well the time 't'. The ions vorticity equation (12) disappears and flow generation mechanism is killed. Equation (10) with $\psi$ defined by Eq. (26) modifies the EMHD result in two ways. One is that the density gradient scale length is not constant and the other is that the generated magnetic field is not unidirectional and constant in space.
In the original Biermann battery effect, the electric field is assumed to be created by the electron pressure gradient, viz,
\begin{equation}
		0=e n_e {\bf E} + \nabla p_e \nonumber \\
\end{equation}
in the background of the stationary ions. The ${\bf v}_e \times {\bf B}$ term in Lorentz force is also neglected. The curl of above equation yields,
\begin{equation}
		\partial_t {\bf B}= -\frac{c}{e}(\frac{\nabla n_e \times \nabla T_e}{n_e})=-\frac{c T_e}{e}(\frac{\nabla n_e}{n_e} \times \frac{\nabla T_e}{T_e}) \nonumber \\
\end{equation}
The density gradient scale length $L_n=\mid \frac{\nabla n_e}{n_e} \mid$ and temperature gradient scale length $L_T=\mid \frac{\nabla T_e}{T_e} \mid$ are assumed to be unidirectional and constant. Thus the generated magnetic field also turns out to be unidirectional perpendicular to both the vectors $\nabla n_e$ and $\nabla T_e$. The simple Biermann battery mechanism or EMHD cannot be applied to the plasma of solar spicules where ions motion is confirmed by observations. To find out an exact two dimensional (2D) analytical solution of the highly nonlinear set of coupled partial differential equations of two fluid plasma including the flows of both species ${\bf v}_j \neq 0$, we have ignored the collisions. Dissipative effects in these physical mechanisms can be included, but in that case the numerical methods are needed to solve the relevant set of equations for galactic plasma and solar atmosphere. It has been shown that the baroclinic vectors can generate flows and magnetic fields simultaneously (Saleem2010, Saleem2021) and ion dynamics should not be ignored.

In the present physical model we have not used the force free condition $(\nabla \times {\bf B})\times {\bf B}=0$ because Eq. (35) implies that, 
\begin{equation}
	(\nabla \times {\bf B}) \times {\bf B} =- (\frac{c T_{e}^{\prime}}{e}R(t))^{2} \eta (\partial_x F, \partial_y F, 0) \neq 0 \nonumber \\
\end{equation}
where 
\begin{equation}
\nabla \times {\bf B}=\frac{c T_{e}^{\prime}}{e} R(t)(0,0,\nabla^2 F)=\frac{c T_{e0}^{\prime}}{e} R(t)(0,0,\eta F)
\end{equation}

\begin{figure}
    \centering
    \includegraphics[width=10cm, height=6cm]{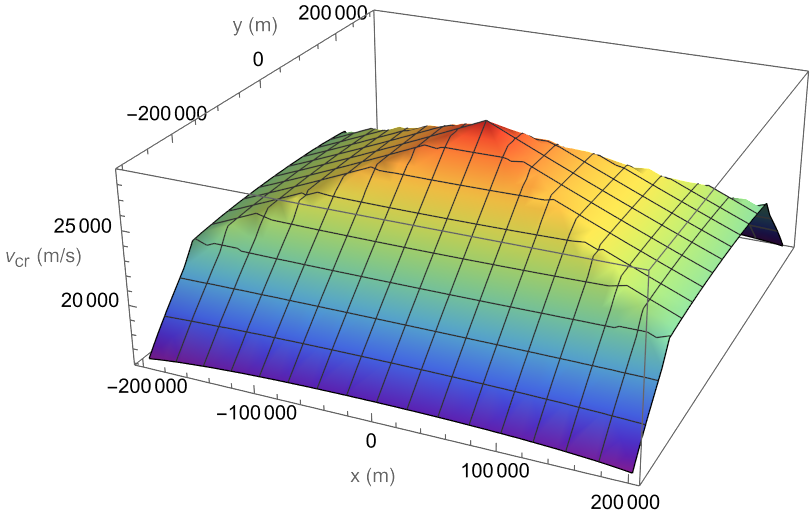}
    \caption{Spatial profile of the upward velocity of ions $[v_i(x,y)]_{ch}$ as function of $(x,y)$ coordinates in the chromosphere. }
    \label{fig4}
\end{figure}

\begin{table}[htbp] % Start the table environment with optional placement options (h for here, t for top, b for bottom, p for page)
    \centering

    \begin{tabular}{|c|c|c|c|c|c|} % Specify the number of columns and their alignment/formatting (c for centered, l for left, r for right, | for vertical lines)
        \hline % Add a horizontal line at the top of the table
        &  \shortstack{\\Time spent by \\density hump $\tau$(s)} & \shortstack{\\ Ion's initial velocity\\  $v_i  (km/s)$ }& \shortstack{\\Ion's final velocity \\$v_f (km/s)$ }&  \shortstack{\\Temperature gradient \\ $T_e^{\prime}$ $(10^{-26}J/m)$}  \\ [4ex] % Table headers
        \hline % Add a horizontal line below the headers
       Chromosphere (ch)& \shortstack{ \\ $\tau_{ch}(0,0)=104 $ \\ \\ $ \tau_{ch}(x_m,y_m)=395$ }& \shortstack{ \\$[v_{i}(0,0)]_{ch}=0 $ \\ \\ $ [v_{i}(x_m,y_m)]_{ch}=0$ } & \shortstack{\\$[v_{f}(0,0)]_{ch}=29 $ \\ \\ $ [v_{f}(x_m,y_m)]_{ch}=8$ } &  19\\ [4ex] % Row 1
        \hline % Add a horizontal line below row 1
      Transition Region (tr)  &\shortstack{\\$\tau_{tr}(0,0)=0.72 $ \\ \\ $ \tau_{tr}(x_m,y_m)=1.18$ }& \shortstack{\\$[v_{i}(0,0)]_{tr}=29 $ \\ \\ $ [v_{i}(x_m,y_m)]_{tr}=8$ }  & \shortstack{\\$[v_{f}(0,0)]_{tr}=108 $ \\  \\$ [v_{f}(x_m,y_m)]_{tr}=76$ }  &3790 \\ [4ex]% Row 2
        \hline % Add a horizontal line below row 2
    Corona (cr) & \shortstack{\\$\tau_{cr}(0,0)=395 $ \\ \\ $ \tau_{cr}(x_m,y_m)=279$ } & \shortstack{\\$[v_{i}(0,0)]_{cr}=108 $ \\ \\ $ [v_{i}(x_m,y_m)]_{cr}=76$ } & \shortstack{\\$[v_{f}(0,0)]_{cr}=0 $ \\ \\ $ [v_{f}(x_m,y_m)]_{cr}=0$ }&  0\\ [4ex] % Row 2
        \hline % Add a horizontal line below row 2
    \end{tabular}
     \caption{The time spent by the density hump $\tau$ and the initial(final) velocities of ions at two points $(0,0)$ and $(x_m,y_m)$ are mentioned in three different regions to show the life cycle of the spicules from birth in the chromosphere with initial velocity zero to death in the corona with final velocity zero. The upward velocities are created by the thermal force propotional to $\nabla n \times (\nabla T_i \times \nabla T_e).$ }
\end{table}

\section{Discussion}
A physical mechanism has been proposed for the generation of spicules and their three stage life cycle. There are two interesting questions related to this phenomenon; one is how the density structure is created in chromosphere with non-zero vertical velocity and the second is why this density structure decays in corona after getting a certain height. 

The process of taking birth of spicules at the bottom of chromosphere has been considered as a sudden rise in local plasma density within a finite area as a function of $(x,y)$ coordinates at time $t=0$ in the presence of gradients of the temperatures of electrons and ions along vertical diretcion i.e. along the z-axis.  
Since temperatures of charged particles have positive gradients along z-direction in chromosphere $\nabla T_j= T_{j0}^{\prime} \hat{z}$, therefore a net upward acceleration is produced by the thermal force which is proportional to the cross product of gradients of density and temperatures $\nabla n(x,y,t) \times (\nabla T_e + \nabla T_i)$. The density structure attains a final non-zero velocity at top of chromosphere after spending time $\tau_{ch}$ here and enters into transition region (TR). The steeper gradients of temperatures along vertical direction in TR give larger upward push to this density structure. The density hump passes through this thin layer of about $100$ $km$ thick in a very short time $\tau_{tr}$. After gaining large velocity it enters into corona where the upward force vanishes due to $\nabla T_j=0$ and the solar gravitational force acting in the negative direction reduces its final upward velocity to zero after time $\tau_{cr}$. The process of the creation of density structures at the bottom of chromosphere continues till the first such structure reaches the point of death in corona with zero final velocity. At this moment, the process of creation of spicule stops. The life time of the spicule is the sum of its time durations spent in the three regions $\tau_l= \tau_{ch} + \tau_{tr} + \tau_{cr}$.

The above model for the genertaion of spicules and their life cycle is based on the two fluid plasma theory. The two fluid plasma equations have been extensivley used to study the dynamics of astrophysical and laboratory plasmas. But it is not straightforward to obtain the exact solutions of this coupled set of highly nonlinear partial differential equations. However, during the past two decades, the two dimensional (2D) \cite{saleem2007theory, saleem2010nonequilibrium, saleem2021exact} and three dimensional (3D) \cite{saleem20223d} exact analytical solutions of these equations have been obtained for a restricted type of flow which is longitudinally uniform with $\nabla \cdot {\bf v}_j=0$ assuming the gradients of density and temperatures to be time-independent $\partial_t n=0$. Recently \cite{saleem20223d}, it has been shown that the jet-like vertical flows can be generated at the surface of neutral fluid or plasma by the thermal force proportional to $\nabla n(x,y) \times (\nabla T_e + \nabla T_i)$ where $\nabla T_j=T_{j0}^{\prime} \hat{z}$. This solution can be applied to the transition region and corona during the life cycle of the spicule. But to explain the generation of the spicule at bottom of the chromosphere, we have to assume that density structure is created during a finite time however small it may be. For this purpose, we assume that the local density hump is created with a delta function in time $\delta(t)$ in the chromosphere. Luckily, the exact solution of two fluid plasma equations turns out to be similar to the one found for the case of constant density with respect to time obtained previously. This new 2D analytical solution with time-dependent density is very interesting because it can explain the process of creation of plasma structures of different sizes and different scales in astrophysical systems. The density function becomes a step function in time and continuity equations have a source term $S(x,y,t)$ for $\partial_t n\neq 0$. It justifies the generation mechanism with consistent mathematical formalism. At present, we have applied this solution to explain the creation of spicules in chromosphere. 

The spatial profile of density hump created in the chromosphere is similar to the function $F(x,y)$ as shown in Fig. (\ref{fig1}).  The $N_0$ is a free parameter and its value is assumed to be larger than the background density i.e. $n_0<N_0$. Finite region of chromosphere has been chosen in the first quadrant to show that in the presence of observed vertical gradients of temperatures in choromosphere and transition region,  all points within this finite density structure i.e $0<x_m, y_m$ get a net upward acceleration and move in vertical direction. However, the accelerations and velocities of different points $(x,y)$ have different values. The life time of the spicule turns out to be a few minutes which is of the order of observed values. But the estimated heights turn out to be larger than the observed ones. The observed heights mentioned in literature are of the order of $10^{6}$ $m$ in general while we obtain the values of the order of $10^{7}$ $m$. If the dissipative effects are taken into account, then these values may reduce. However, it should be noted that the combined thickness of chromosphere and  transition region is $(1.6) \times 10^{6}$ $m$ in our model and we have included the total length of the structure from foot of the spicule in chromosphere to the head in corona. It may happen that by the time, it reaches its final destiny in corona, the lower parts had vanish already. On the other hand, since the upward flow is not uniform, therefore the different layers of flow within the structure give rise to resisitive forces which have not been taken into account in the present analytical model based on exact solutions of two fluid equations. The variation of density with respect to altitude may also cause modifications in the shape and acceleration etc. Such considerations are beyond the scope of the present work.

The thermal force mainly creates the ions vorticity $(\nabla \times {\bf v}_i)$ which is coupled with the generation of magnetic field. The generated vorticity requires the ions to flow in the vertical direction, viz. ${\bf v}_i= v_{i} (x,y,t) \hat{z}$. The ion velocity is related to the flow of electrons through Ampere's law, and hence the plasma jet-like flow along the z-axis is created. The magnitude of the generated magnetic field ${\bf B}(x,y,t)$ turns out to be very small compared to the constant ambient magnetic field  ${\bf B}_0$ because the scale lengths of gradients are large. 
The Figs. (\ref{fig2}) and (\ref{fig3}) represent the behavior of physical quantities, but are not according to physical scales. The Fig. (\ref{fig4}) shows the variation of vertical velocity within the spicule as a function of $(x,y)$ coordinates. This is a general model which can explain the fundamental physical mechanism for the generation of vertical plasma motions of different spatial scales in the solar atmosphere. It can be applied to the generation of several different type of structures such as threads and strands within coronal loops, solar mass ejection and it also seems to add plasma in the solar wind at least partially.

Acknowledgement\\
Authors would like to thank the anonymous referee for his very valuable suggestions and comments which have helped the authors in making reasonable numerical estimates and improving the presentation of this research work.
\bibliography{bibliography}{}
\bibliographystyle{aasjournal}

\end{document}